\documentclass[doublecol]{epl2} 
\usepackage{amsmath,bm,graphicx}

\title{Tagged particle in a sheared suspension: effective temperature 
determines density distribution in a slowly varying external potential
beyond linear response}
\shorttitle{Effective temperature beyond linear response}

\author{Grzegorz Szamel and Min Zhang}
\shortauthor{G. Szamel and M. Zhang}

\institute{Department of Chemistry, 
Colorado State University, Fort Collins, CO 80523
}

\date{\today}


\abstract{
We consider a sheared colloidal suspension under the influence of an
external potential that varies slowly in space in the plane perpendicular to
the flow and acts on one selected (tagged) particle of the 
suspension. Using a Chapman-Enskog type expansion we 
derive a steady state equation for the tagged particle
density distribution. We show that for potentials varying along
one direction only, the tagged particle distribution is the
same as the equilibrium distribution with the temperature equal to
the effective temperature obtained from the violation of the Einstein
relation between the self-diffusion and tagged particle mobility coefficients.
We thus prove the usefulness of this effective temperature
for the description of the tagged particle behavior beyond the realm of linear 
response. We illustrate our theoretical predictions with Brownian dynamics computer
simulations. 
}

\begin{document}

\maketitle

\section{Introduction} 

While the principles of equilibrium statistical 
mechanics are well established, a general framework of its non-equilibrium
counterpart is still lacking. Recently, there has been a lot of interest 
in various facets of non-equilibrium statistical mechanics
\cite{nesm}. One of the recurring concepts is the so-called effective temperature
\cite{LCreview}. Originally, it was defined through
the violation of the fluctuation-dissipation relation in mean-field
spin systems \cite{CKP}. Subsequently, a variety of definitions of the 
effective temperature, in terms of violations of different dynamic and static linear
response type relations, have been investigated using computer simulations
of model fluids \cite{LCreview,BerBar,Liu,LSS,ZS}. Notably, there have been few
theoretical studies of the effective temperature \cite{HS,SS,Szamel,KF}. 

The original definition of the effective temperature 
and almost all subsequent investigations mentioned above 
were concerned with the linear response regime. 
On the other hand, in equilibrium statistical mechanics the 
temperature plays an important role also outside the realm of linear response. 
Hayashi and Sasa \cite{HS} were the first to discuss an application of the effective 
temperature outside the linear response regime. They studied an 
exactly solvable model, a single Brownian particle in a tilted washboard potential. 
They showed that the effective temperature defined through the violation
of the Einstein relation determines the density distribution in a long-wavelength
external potential. Subsequently, there were three simulational studies of the
effective temperature outside the linear response regime. Importantly, these studies
dealt with strongly interacting many-body systems. Zamponi \textit{et al.} \cite{ZRA} 
showed that the effective temperature enters into the fluctuation relation 
\cite{fluctrel} for a driven glassy fluid. Ilg and Barrat \cite{IB} showed that 
the effective temperature determines the barrier crossing rate in a driven glassy 
system. Haxton and Liu \cite{HL} investigated
the usefulness of the effective temperature for the description of material
properties (specifically, the shear stress and the average inherent structure energy) 
of driven glassy systems. To date, there is no theoretical analysis of the 
effective temperature of a many-body system, outside the linear response regime.

Establishing the usefulness of the effective temperature outside the linear
response regime is essential because, in principle, 
the violation of the fluctuation-dissipation relation does not have to be
interpreted in terms of the effective temperature. The best illustration
of this fact is the difference between two studies of the 
same exactly solvable model, a single Brownian particle in a tilted washboard potential. 
Hayashi and Sasa \cite{HS} used the violation of the fluctuation-dissipation relation
to define the effective temperature. On the other hand, Speck and Seifert \cite{SS}  
showed that the fluctuation-dissipation relation can be effectively restored 
if fluctuations of the Brownian particle velocity are measured with respect to the local 
mean velocity. We should also mention a recent investigation \cite{BMW} which
strives to provide a unifying framework for non-equilibrium fluctuation-dissipation 
relations. It is similar to Speck and Seifert's approach in that it focuses on an
additive correction responsible for the  violation of the fluctuation-dissipation 
relation. It de-emphasizes the effective temperature altogether \cite{SS2}. 
These examples show the need for additional, fundamental understanding of the meaning
of the effective temperature, especially beyond the realm of linear response. 

In this Letter we show theoretically that for an interacting many-body system 
one specific effective temperature 
plays the role of the usual temperature outside the linear response regime.
We consider a sheared colloidal suspension. 
In such a suspension the Einstein relation between the self-diffusion 
coefficient and the tagged particle mobility is violated \cite{SBL} and 
this violation can be used to define the effective temperature 
\cite{BerBar,Szamel,KF}. We assume that the suspension is under the
influence of a spatially varying external potential which changes 
slowly along one direction in the plane perpendicular to the flow and acts only on 
one selected (tagged) particle of the suspension. 
We show that the functional form of the tagged particle density 
distribution is then the same as that of the equilibrium distribution. However, 
in this distribution the temperature is replaced by 
the effective temperature obtained from the violation of
the Einstein relation along the same direction. Consequently, 
we can also prove that the same effective temperature is obtained from the 
tagged particle static linear response relation and the Einstein relation violation.

Since the derivation of the main result is a little technical, we first
present the results and illustrate them with Brownian dynamics computer
simulations. We sketch the derivation in the latter part of this Letter.

\section{Tagged particle density distribution} 
 
The $N$-particle steady state probability distribution $P_\text{s}^V$
describing a suspension undergoing a shear flow and under an influence of
an external potential satisfies the following equation,
\begin{equation}\label{SmolV}
\left[\Omega_{\text{s}} +  \partial_{\mathbf{r}_1} \cdot 
\mu_0 \left(\partial_{\mathbf{r}_1} V^{\text{ext}}(\mathbf{r}_1)\right) \right]
P_{\text{s}}^V(\mathbf{r}_1, ..., \mathbf{r}_N) = 0,
\end{equation}
where $\Omega_{\text{s}}$ is the Smoluchowski operator with shear flow,
\begin{equation}\label{SmolO}
\Omega_{\text{s}} = \sum_{i=1}^N \partial_{\mathbf{r}_i} \cdot
\left( D_0 \partial_{\mathbf{r}_i} -  
\mu_0 \sum_{i\neq j =1}^N \mathbf{F}(\mathbf{r}_{ij})
- \mathbf{v}(\mathbf{r}_i) \right).
\end{equation}
In the above equations, $\mu_0$ and $D_0$ is the mobility and the diffusion
coefficient, respectively, of an isolated Brownian particle, 
$\mathbf{F}(\mathbf{r}_{ij})$ is the force acting on particle $i$ due to particle $j$,
$\mathbf{r}_{ij} = \mathbf{r}_i - \mathbf{r}_j$, 
and $\mathbf{v}$ is the 
uniform shear flow, $\mathbf{v}(\mathbf{r}_i) = \mathbf{\Gamma}\cdot\mathbf{r}_i$, 
with $\mathbf{\Gamma}$ being the velocity gradient tensor. 
Finally, $V^{\text{ext}}$ is an external potential. It acts only on 
the tagged particle and varies in the plane perpendicular to the
flow, $\partial_{\mathbf{r}_1}\cdot\mathbf{\Gamma} V^{\text{ext}}(\mathbf{r}_1) = 0$.

We should emphasize that Eqs. (\ref{SmolV}-\ref{SmolO}) without the external potential 
have been widely used in investigations of the dynamics and rheology of sheared
suspensions \cite{FC}. There are two assumptions implicit in writing down these
equations. First, hydrodynamic interactions are neglected. Second, 
it is assumed that the shear rate is large enough to drive Brownian
particles out of equilibrium, but it is small enough so that the solvent
in which these particles are suspended is in equilibrium. The latter assumption implies 
that the Einstein relation for an isolated particle is satisfied, 
$D_0=k_B T \mu_0$, where $T$ is the solvent temperature.

In principle, in order to find the steady state 
tagged particle distribution $n_{\text{s}}$ we need to solve 
Eq. \eqref{SmolV} and then integrate over the positions of all other particles,
\begin{equation}\label{nsfromPsV}
n_{\text{s}}(\mathbf{r}_1) = \int d\mathbf{r}_2 ... d\mathbf{r}_N
P_{\text{s}}^V(\mathbf{r}_1, ..., \mathbf{r}_N).
\end{equation}
In the absence of the shear flow, we know the solution of Eq. \eqref{SmolV}  
and we can easily find the equilibrium distribution for the tagged particle in 
external potential $V^{\text{ext}}$,
\begin{equation}\label{nVeq}
n(\mathbf{r}_1) \propto \exp(-V^{\text{ext}}(\mathbf{r}_1)/(k_B T)).
\end{equation}
In the latter part of this Letter we use a Chapman-Enskog type expansion to 
show that, in the presence of the shear flow,
for an external potential which varies slowly in the plane perpendicular
to the flow, the steady state tagged particle distribution 
satisfies the following equation,
\begin{equation}\label{nssseq}
\partial_{\mathbf{r}_{\perp}}\cdot
\left( \mathbf{D} \cdot \partial_{\mathbf{r}_{\perp}}
+  \bm{\mu} \cdot 
\left(\partial_{\mathbf{r}_{\perp}} V^{\text{ext}}(\mathbf{r}_{\perp})\right) \right) 
n_{\text{s}}(\mathbf{r}_{\perp}) = 0.
\end{equation}
Here $\mathbf{r}_{\perp}$ is a two-dimensional vector in the plane perpendicular
to the flow, and $\mathbf{D}$ and $\bm{\mu}$
are the self-diffusion and the tagged particle mobility tensors which depend on 
the suspension's density, temperature, and shear rate.

It follows from Eq. \eqref{nssseq} that, if external potential $V^{\text{ext}}$
varies only in the direction of unit vector $\hat{\mathbf{n}}$ (in the plane
perpendicular to the flow), the tagged particle
distribution has the form of the equilibrium distribution \eqref{nVeq},
\begin{equation}\label{nVss}
n_{\text{s}}(\mathbf{r}_{\perp}\cdot\hat{\mathbf{n}}) \propto
\exp(-V^{\text{ext}}(\mathbf{r}_{\perp}\cdot\hat{\mathbf{n}})
/(k_B T_{\hat{\mathbf{n}}}^{\text{eff}})),
\end{equation}
but with the temperature being replaced by the effective temperature defined 
in terms of the violation 
of the Einstein relation along direction $\hat{\mathbf{n}}$,
\begin{equation}\label{Teff}
T_{\hat{\mathbf{n}}}^{\text{eff}} = 
\frac{\hat{\mathbf{n}}\cdot\mathbf{D}\cdot \hat{\mathbf{n}}}
{\hat{\mathbf{n}}\cdot\bm{\mu}\cdot \hat{\mathbf{n}}}.
\end{equation}
For large enough shear rates the system can become structurally anisotropic 
even though the density is still uniform. In particular, structure factors 
along different directions in the reciprocal space vary differently with
the shear rate \cite{ZS}. In principle, one cannot expect that in such cases
the effective temperature is isotropic. On the other hand, for a system with
very slow dynamics (\textit{e.g.} for a glassy system) there is a range of shear
rates for which the system is strongly out of equilibrium but still structurally 
isotropic, with essentially shear-rate-independent structure factors along all 
directions in the reciprocal space \cite{BerBar}. In such cases it has been found
that the effective temperature is isotropic in the linear response regime\cite{BerBar}.
We expect (although we did not investigate it) that the same will be true 
also beyond the realm of the linear response. Of course, for any ergodic system  
$T^{\text{eff}}\to T$ in the limit $\dot{\gamma}\to 0$.

One of the consequences of Eq. \eqref{nVss} is the following relation between 
an infinitesimally small external potential $\delta V^{\text{ext}}$ and
the resulting change $\delta n_{\text{s}}$ of the tagged particle distribution,
\begin{equation}\label{linres}
\delta n_{\text{s}}(\mathbf{r}_{\perp}\cdot\hat{\mathbf{n}})
= - \frac{ n_{\text{s}}}
{k_B T_{\hat{\mathbf{n}}}^{\text{eff}}} \; \delta 
V^{\text{ext}}(\mathbf{r}_{\perp}\cdot\hat{\mathbf{n}}).
\end{equation}
Eq. \eqref{linres} is a generalization of the familiar equilibrium linear
response relation. It is analogous to static linear response type relations 
that were used to define effective temperatures by A. Liu and collaborators \cite{Liu}.
Importantly, in Eq. \eqref{linres} effective temperature \eqref{Teff} appears
in the place of a static linear response-based effective temperature. 
This is the first time that the effective temperature defined in terms
of the violation of a dynamic fluctuation-dissipation relation 
has been shown theoretically to be identical to the effective temperature defined in
terms of the violation of a static linear response relation.

\section{Brownian dynamics simulations} 

We performed Brownian dynamics computer simulations as described in Ref.~\cite{ZS}.  
The system consists of $N=1372$ particles interacting via
a screened Coulomb potential,
\begin{equation}
V(r) = A \exp\left(-\kappa(r-\sigma)\right)/r,
\end{equation}
with $A=475 k_B T \sigma$ and $\kappa\sigma=24$.
We simulated the system at a dimensionless density,
$N\sigma^3/L^3$, equal to 0.408 ($L$ is the box length), which
corresponds to the hard-sphere volume fraction equal to 0.43 \cite{GS}.
We applied shear flow in the $x$ direction with the velocity
gradient in the $y$ direction, 
$\mathbf{\Gamma} = \dot{\gamma}\hat{\mathbf{x}}\hat{\mathbf{y}}$ 
with 
$\dot{\gamma}$ being the shear rate and $\hat{\bm{\alpha}}$ the unit
vector along the $\alpha$ axis.
%, $\alpha=x,y$. 
In the following we use reduced units: we measure lengths in units of $\sigma$,
time in units of $\sigma^2/D_0$, energy in units of $k_B T$, and effective temperature
in units of $T$.

\begin{figure}
\includegraphics[width=3.4in,clip=true]{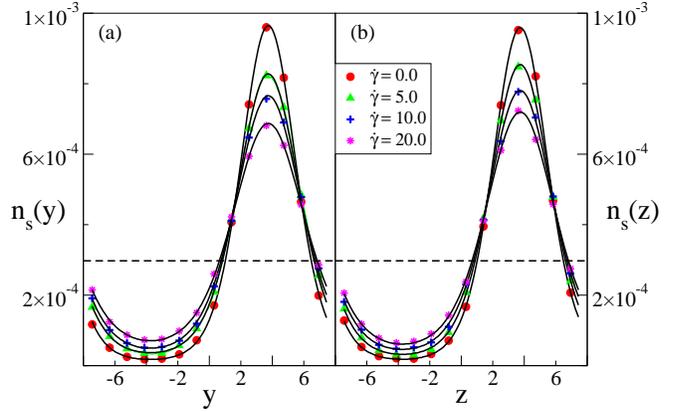}
\caption{\label{fig1} (Color online) 
Tagged particle steady state density distribution for suspensions 
with the external potential varying along (a) velocity gradient direction and (b) 
vorticity direction. Symbols: Brownian dynamics simulations' results.
Solid lines: fits to a normalized version of Eq. \eqref{nVss}.
Dashed line: tagged particle density in the absence of
the external potential.}
\end{figure}

\begin{figure}
\includegraphics[width=3.4in,clip=true]{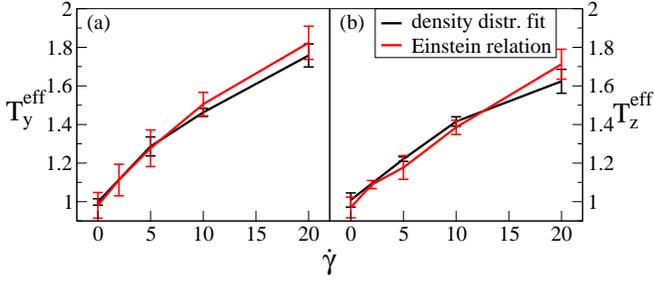}
\caption{\label{fig2} (Color online) 
Comparison of shear rate dependence of effective temperatures 
obtained from fits of tagged particle steady state density distributions to 
Eq. \eqref{nVss} and from the violation of the Einstein relation. 
(a) velocity gradient direction, $\hat{\mathbf{y}}$. 
(b) vorticity direction, $\hat{\mathbf{z}}$.
}
\end{figure}

To illustrate the validity of Eq. \eqref{nVss} we introduced 
a slowly varying external potential acting on particle number 1 
\cite{comment1}. We chose the simplest, plane wave potential, 
\begin{equation}
V^{\text{ext}}(\alpha_1) = V_0 \sin(2\pi \alpha_1/L),
\end{equation} 
where $\alpha=y,z$. Note that $2\pi/L$ is the smallest wave-vector
allowed by the periodic boundary conditions. We used $V_0=2$, which is 
well outside the linear response regime (see Fig. \ref{fig1}). 
For each choice of the potential
we performed 4 independent runs at the following shear rates: 0, 5, 10, and 20. 
We monitored the tagged particle density distribution, 
$n_{\text{s}}(\alpha)$, $\alpha=y,z$. All results presented 
are averaged over 4 runs. 

In Fig. \ref{fig1} we show tagged particle steady state density distributions.
Qualitatively, with increasing shear rate the influence of the
external potential decreases, which could be interpreted as an increasing effective
temperature. More importantly, the simulations' results can
be fitted very well using equilibrium-like distributions \eqref{nVss} with 
%$T_{\hat{\mathbf{n}}}^{\text{eff}}$ 
an effective temperature being the only fit parameter. 
In Fig. \ref{fig2} we show that 
the values of $T_{\hat{\bm{\alpha}}}^{\text{eff}}$,  
$\hat{\bm{\alpha}}=\hat{\mathbf{y}}, \hat{\mathbf{z}}$,
obtained from the fits agree very well with 
the previously obtained effective temperatures defined through the violation 
of the Einstein relation \cite{ZS}. 

At the largest shear rates
we found previously that the system is structurally anisotropic \cite{ZS}. 
Thus, in principle, different effective temperatures can be obtained along different 
directions. We note that results showed in Fig. \ref{fig2} suggest slightly
different effective temperatures along $\hat{\mathbf{y}}$ and $\hat{\mathbf{z}}$
directions. However, the difference between these temperatures 
is within error bars of our simulations.

\section{Derivation} 
 
We make two assumptions about the external potential acting on the 
tagged particle. First, we assume that it is transverse, 
\textit{i.e.} that it varies only in the plane perpendicular to the flow, 
$\partial_{\mathbf{r}_1}\cdot\mathbf{\Gamma} V^{\text{ext}}(\mathbf{r}_1) = 0$. 
Second, we assume that the potential is slowly varying. To make the latter 
assumption explicit and to facilitate the subsequent gradient expansion
we introduce a small parameter $\epsilon$ and write the external
potential as $V^{\text{ext}}(\epsilon\mathbf{r}_1)$. As usual, 
at the end of the calculation we will put $\epsilon=1$. We emphasize that 
no assumption is made regarding the strength of the external potential.

It is natural to assume that for a slowly varying, transverse external potential
the tagged particle density distribution $n_{\text{s}}$ will also be slowly varying 
and transverse. To find a steady state equation for $n_{\text{s}}$ we use the
Chapman-Enskog \cite{Resibois} expansion. Specifically, we follow Titulaer 
\cite{Titulaer} who applied the Chapman-Enskog expansion to the problem
of deriving the so-called contracted, Smoluchowski description of a Brownian
particle in position space from the full Fokker-Planck description in velocity and
position space. We note that for \textit{linear} equations considered in Ref. 
\cite{Titulaer} and in our Letter, the Champan-Enskog procedure is closely 
related to a variant of perturbation theory \cite{Titulaer2}.

Before applying the Chapman-Enskog expansion, 
we have to face the fact that, although the tagged particle density $n_{\text{s}}$ is 
slowly varying, due to the inter-particle interactions $N$-particle distribution 
$P^V_{\text{s}}$ is \textit{not} a slowly varying function of the 
tagged particle position $\mathbf{r}_1$.
To disentangle the slow variation of $P^V_{\text{s}}$ with $\mathbf{r}_1$, which is 
induced by the external potential, and the rapid variation 
$P^V_{\text{s}}$ with $\mathbf{r}_1$, which 
originates from the inter-particle interactions, we introduce
new variables: $\mathbf{R}_1 = \epsilon \mathbf{r}_1, \mathbf{R}_2 = \mathbf{r}_{21},
...,  \mathbf{R}_N = \mathbf{r}_{N1}$. In these new variables the steady
state equation \eqref{SmolV} has the following form:
\begin{eqnarray}\label{SmolVscaled}
\left[ \Omega^{(0)} + \epsilon\Omega^{(1)} + \epsilon^2 \Omega^{(2)} \right]
P_{\text{s}}^V(\mathbf{R}_1, ..., \mathbf{R}_N) = 0.
\end{eqnarray}
In Eq. \eqref{SmolVscaled},
\begin{eqnarray}\label{SmolVscaled0}
\Omega^{(0)} \!\!\!\!\! &=&  \!\!\!\!\!
- \partial_{\mathbf{R}_1}\cdot\mathbf{\Gamma}\cdot\mathbf{R}_1 + 
\sum_{i=2}^N \partial_{\mathbf{R}_i} \cdot
\left( D_0 \partial_{\mathbf{R}_i} + D_0\sum_{j=2}^N \partial_{\mathbf{R}_j} 
\right. \nonumber \\ && \!\!\!\!\! \left. 
- \mu_0 \sum_{i\neq j=2}^N \mathbf{F}(\mathbf{R}_{ij}) + 
\mu_0 \sum_{j=2}^N \mathbf{F}(-\mathbf{R}_j)
-  \mathbf{\Gamma}\cdot\mathbf{R}_i 
\right)
\\[1ex] \label{SmolVscaled1}
\Omega^{(1)} \!\!\!\!\! &=&  \!\!\!\!\!
-\partial_{\mathbf{R}_1}\cdot \left( D_0\sum_{i=2}^N \partial_{\mathbf{R}_i}
+ \mu_0 \sum_{i=2}^N \mathbf{F}(-\mathbf{R}_i) \right)
\\ \nonumber && \!\!\!\!\!
- \sum_{i=2}^N \partial_{\mathbf{R}_i} \cdot \mu_0
\left( \partial_{\mathbf{R}_1}  V^{\text{ext}}(\mathbf{R}_1) \right)
- 
D_0 \sum_{i=2}^N \partial_{\mathbf{R}_i} \cdot \partial_{\mathbf{R}_1}
\\[1ex] \label{SmolVscaled2}
\Omega^{(2)} \!\!\!\!\! &=&  \!\!\!\!\!
\partial_{\mathbf{R}_1}\cdot\left( 
D_0\partial_{\mathbf{R}_1}
+ \left( \partial_{\mathbf{R}_1}  V^{\text{ext}}(\mathbf{R}_1) \right) \right).
\end{eqnarray}

We now follow Titulaer \cite{Titulaer} and simultaneously find a special solution
of Eq. \eqref{SmolVscaled} 
\begin{eqnarray}\label{Pexp}
&& P_{\text{s}}^V(\mathbf{R}_1, ..., \mathbf{R}_N) =  
n_{\text{s}}(\mathbf{R}_1) 
P_{\text{s}}^{(0)}(\mathbf{R}_2, ..., \mathbf{R}_N) 
\\ \nonumber && +
\epsilon P_{\text{s}}^{(1)}(\mathbf{R}_1,\mathbf{R}_2, ..., \mathbf{R}_N) +
\epsilon^2 P_{\text{s}}^{(2)}(\mathbf{R}_1,\mathbf{R}_2, ..., \mathbf{R}_N) + ...
\end{eqnarray}
and a steady state equation for the tagged particle
distribution,
\begin{eqnarray}\label{nseqexp}
\left( \mathcal{D}^{(0)} + \epsilon \mathcal{D}^{(1)} + 
\epsilon^2 \mathcal{D}^{(2)} + ...
\right) n_{\text{s}}(\mathbf{R}_1)  = 0. 
\end{eqnarray}
Here Eq. \eqref{nseqexp} is obtained from integrating Eq. \eqref{SmolVscaled} 
over $\mathbf{R}_i$, $i>1$. Thus, \textit{e.g.},
\begin{eqnarray}\label{nseqexp0}
\mathcal{D}^{(0)} n_{\text{s}}(\mathbf{R}_1) \!\!\!\! &=& \!\!\!\!
\int d\mathbf{R}_2 ... d\mathbf{R}_N \Omega^{(0)} n_{\text{s}}(\mathbf{R}_1) 
P_{\text{s}}^{(0)}(\mathbf{R}_2, ..., \mathbf{R}_N),
\nonumber \\
\end{eqnarray}
and
\begin{eqnarray}\label{nseqexp1}
\mathcal{D}^{(1)} n_{\text{s}}(\mathbf{R}_1) \!\!\!\! &=& \!\!\!\!
\int d\mathbf{R}_2 ... d\mathbf{R}_N \Omega^{(1)} n_{\text{s}}(\mathbf{R}_1) 
P_{\text{s}}^{(0)}(\mathbf{R}_2, ..., \mathbf{R}_N)
\nonumber \\
\!\!\!\! &+& \!\!\!\! 
\int d\mathbf{R}_2 ... d\mathbf{R}_N \Omega^{(0)} 
P_{\text{s}}^{(1)}(\mathbf{R}_1,\mathbf{R}_2, ..., \mathbf{R}_N).
\nonumber \\
\end{eqnarray}
It should be emphasized that in the Chapman-Enskog
type procedure the steady state tagged particle distribution $n_{\text{s}}$
is \textit{not} expanded in $\epsilon$. Moreover, 
higher order terms $P_{\text{s}}^{(i)}$, $i>0$, are determined up to 
an arbitrary multiple of $P_{\text{s}}^{(0)}$ \cite{Titulaer}. 
We follow the usual procedure \cite{Titulaer} and impose conditions
\begin{equation}
\int d\mathbf{R}_2 ... d\mathbf{R}_N 
P_{\text{s}}^{(i)}(\mathbf{R}_1, ..., \mathbf{R}_N) = 0 \text{ for } i>0,
\end{equation}
which fix these contributions to be zero. It should be noted that due to these
conditions the tagged particle distribution $n_{\text{s}}$
is completely determined by the first term in expansion \eqref{Pexp}.

We substitute \eqref{Pexp} into steady state equation \eqref{SmolVscaled},
solve iteratively for $P_{\text{s}}^{(i)}$, and calculate
$\mathcal{D}^{(i)}$. The results of this procedure are as follows. First,
$P_{\text{s}}^{(0)}$ is the translationally invariant solution of the 
Smoluchowski equation with shear flow in the absence of an external potential.
This solution depends only on $\mathbf{R}_i$, $i>1$,  
and is normalized in the following way:
$\int d\mathbf{R}_2 ... d\mathbf{R}_N P_{\text{s}}^{(0)} =1$.
Second, 
$\mathcal{D}^{(0)}=\partial_{\mathbf{R}_1}\cdot\mathbf{\Gamma}\cdot\mathbf{R}_1$.
Thus, for $n_{\text{s}}(\mathbf{R}_1)$ that varies only in the plane perpendicular
to the flow $\mathcal{D}^{(0)}$ 
does not contribute to steady state equation \eqref{nseqexp}.
Third, $\mathcal{D}^{(1)}$ vanishes. Fourth, $\mathcal{D}^{(2)}$ is determined
by the following equation
\begin{eqnarray}\label{eps2}
\lefteqn{\mathcal{D}^{(2)} n(\mathbf{R}_1) = }
\\ \nonumber &&  
\left[ D_0 \partial_{\mathbf{R}_1}\cdot\partial_{\mathbf{R}_1} + 
\partial_{\mathbf{R}_1}\cdot
\mu_0 \left( \partial_{\mathbf{R}_1} V^{\text{ext}}(\mathbf{R}_1) \right) \right]
n(\mathbf{R}_1)
\\ \nonumber &&  
- \partial_{\mathbf{R}_1} \! \cdot \! \int d\mathbf{R}_2 ... d\mathbf{R}_N
\mu_0 \sum_{i=2}^N \mathbf{F}(-\mathbf{R}_i)  
P_{\text{s}}^{(1)}(\mathbf{R}_1, ..., \mathbf{R}_N).
\end{eqnarray}
To complete the calculation, we determine $P_{\text{s}}^{(1)}$ using terms of 
order $\epsilon$ in Eq. \eqref{SmolVscaled},
\begin{eqnarray}\label{P1}
\Omega^{(0)} P_{\text{s}}^{(1)}(\mathbf{R}_1, ..., \mathbf{R}_N) 
= - \Omega^{(1)} n_{\text{s}}(\mathbf{R}_1) 
P_{\text{s}}^{(0)}(\mathbf{R}_2, ..., \mathbf{R}_N).
\end{eqnarray}
Using the explicit form of $\Omega^{(1)}$, Eq. \eqref{SmolVscaled1}, it can be showed
that the source at the right-hand-side of Eq. \eqref{P1} consists of 
two terms, which are proportional to 
$\left[\partial_{\mathbf{R}_1}n_{\text{s}}(\mathbf{R}_1)\right]$ and 
$\left[ \partial_{\mathbf{R}_1}  V^{\text{ext}}(\mathbf{R}_1) \right]
n_{\text{s}}(\mathbf{R}_1)$:
\begin{eqnarray}\label{sourcea}
&& - \Omega^{(1)} n_{\text{s}}(\mathbf{R}_1) 
P_{\text{s}}^{(0)}(\mathbf{R}_2, ..., \mathbf{R}_N)
\nonumber \\ && = \left[\partial_{\mathbf{R}_1}n_{\text{s}}(\mathbf{R}_1)\right]
\left(2 D_0 \sum_{i=2}^N \partial_{\mathbf{R}_i}
+ \mu_0 \sum_{j\neq 1} \mathbf{F}(-\mathbf{R}_j) \right)  
\nonumber \\ && \times P_{\text{s}}^{(0)}(\mathbf{R}_2, ..., \mathbf{R}_N)
\nonumber \\ && + 
\left[ \partial_{\mathbf{R}_1}  V^{\text{ext}}(\mathbf{R}_1) \right]
n_{\text{s}}(\mathbf{R}_1) \mu_0  \sum_{i=2}^N \partial_{\mathbf{R}_i} 
%\nonumber \\ && \times 
P_{\text{s}}^{(0)}(\mathbf{R}_2, ..., \mathbf{R}_N).
\nonumber \\
\end{eqnarray}
To get $P_{\text{s}}^{(1)}$ we now have to solve Eq. \eqref{P1}. While inverting
operator $\Omega^{(0)}$ we note that since both $n_{\text{s}}(\mathbf{R}_1)$
and $V^{\text{ext}}(\mathbf{R}_1)$ vary only in the plane perpendicular
to the flow, operator $\left[\Omega^{(0)}\right]^{-1}$ will not act on them.
Therefore, $P_{\text{s}}^{(1)}$ 
also consists of two terms which are proportional to 
$\left[\partial_{\mathbf{R}_1}n_{\text{s}}(\mathbf{R}_1)\right]$ and 
$\left[ \partial_{\mathbf{R}_1}  V^{\text{ext}}(\mathbf{R}_1) \right]
n_{\text{s}}(\mathbf{R}_1)$.
%\begin{eqnarray}\label{P1a}
%P_{\text{s}}^{(1a)}(\mathbf{R}_1, ..., \mathbf{R}_N) = 
%[\Omega^{(0)}]^{-1} \left( 2 D_0 \sum_{i=2}^N \partial_{\mathbf{R}_i}
%+ \mu_0 \sum_{j\neq 1} \mathbf{F}(-\mathbf{R}_j) \right)
%P_{\text{s}}^{(0)}(\mathbf{R}_2, ..., \mathbf{R}_N) 
%\partial_{\mathbf{R}_1}n_{\text{s}}(\mathbf{R}_1).
%\end{eqnarray}
After these two terms are substituted
into Eq. \eqref{eps2} we can easily derive the following equality
\begin{eqnarray}\label{eps2a}
\lefteqn{\mathcal{D}^{(2)} n(\mathbf{R}_1) = 
\left[\partial_{\mathbf{R}_1} \cdot \mathbf{D} \cdot \partial_{\mathbf{R}_1} \right.
}
\\ \nonumber &&  \left. +
\partial_{\mathbf{R}_1}\cdot \bm{\mu} \cdot
\left( \partial_{\mathbf{R}_1} V^{\text{ext}}(\mathbf{R}_1) \right) \right]
n(\mathbf{R}_1).
\end{eqnarray}
We now show that tensors $\mathbf{D}$ and $\bm{\mu}$, which enter into
Eq. \eqref{eps2a}, are the self-diffusion tensor and the 
tagged particle mobility tensor, respectively. Explicitly, 
\begin{equation}\label{D}
\mathbf{D} = D_0\mathbf{I} + \mathbf{D}^{\text{in}}
\end{equation}
where $\mathbf{I}$ is the 
unit tensor and the interaction contribution $ \mathbf{D}^{\text{in}}$ is 
given by the following formula,
\begin{eqnarray}\label{Dintscaled}
\mathbf{D}^{\text{in}} &=& \int d\mathbf{R}_2 ... d\mathbf{R}_N  
\sum_{i=2}^N \mathbf{F}(-\mathbf{R}_i)
[\Omega^{(0)}]^{-1} 
\nonumber \\ && \times 
\left( - 2 D_0 \sum_{i=2}^N \partial_{\mathbf{R}_i}
- \mu_0 \sum_{j\neq 1} \mathbf{F}(-\mathbf{R}_j) \right)
\nonumber \\ && \times  
P_{\text{s}}^{(0)}(\mathbf{R}_2, ..., \mathbf{R}_N).
\end{eqnarray}
After returning to the original variables 
$\mathbf{r}_1, ..., \mathbf{r}_N$ and putting $\epsilon=1$,  
the interaction contribution $\mathbf{D}^{\text{in}}$ can be 
written in the following form:
\begin{eqnarray}\label{Dint}
\mathbf{D}^{\text{in}} &=& \frac{1}{V}\int d\mathbf{r}_1 ... d\mathbf{r}_N
\mu_0 \mathbf{F}_1 \Omega_s^{-1} 
\left( 2 D_0\partial_{\mathbf{r}_1} - \mu_0 \mathbf{F}_1 \right)
\nonumber \\ && \times 
P_s^{(0)}(\mathbf{r}_1,...,\mathbf{r}_N)
\end{eqnarray}
where $\mathbf{F}_1$ is the force acting on the tagged particle,
$\mathbf{F}_1 = \sum_{i=2}^N \mathbf{F}(\mathbf{r}_{1i})$. 
Following very similar steps
one can show that 
\begin{equation}\label{mu}
\bm{\mu} = \mu_0\mathbf{I} + \bm{\mu}^{\text{in}}
\end{equation}
where 
\begin{eqnarray}\label{muint}
\bm{\mu}^{\text{in}} = \frac{1}{V}\int d\mathbf{r}_1 ... d\mathbf{r}_N
\mu_0 \mathbf{F}_1 \Omega_s^{-1} 
D_0\partial_{\mathbf{r}_1}
P_s^{(0)}(\mathbf{r}_1,...,\mathbf{r}_N).
\end{eqnarray}
The above expressions are identical to the expressions  
for the self-diffusion tensor and the 
tagged particle mobility tensor that have been derived in Refs. \cite{Szamel,KF}.

Finally, we return to the original variables in Eq. \eqref{eps2a}, 
put $\epsilon=1$, use the fact that both the external potential and the tagged
particle density vary only in the plane perpendicular to the flow, and
arrive at Eq. \eqref{nssseq}.

\section{Discussion} 

We showed that the effective temperature 
defined through the violation of the Einstein relation determines the tagged
particle density distribution induced by a slowly varying external potential 
acting on the tagged particle. As a consequence, the same effective temperature
enters into the violation of the Einstein relation and the modified static
linear response relation for the tagged particle. 
It would be very interesting
to extend both results to at least some collective properties. This would
provide additional fundamental justification for introducing the effective temperature. 
It would also allow some
deeper theoretical understanding of the very intriguing observation made by Liu 
and collaborators \cite{Liu} that for \textit{some} combinations of variables
one should use dynamic linear response relations to determine the effective
temperature whereas for \textit{other} combinations of variables one could
use static linear response relations.

We note in this context that the most obvious (and perhaps the most
naive) generalization of the present results to the collective density is not valid
already in the linear response regime: 
we showed earlier \cite{ZS} that the effective
temperature obtained from the collective version of the modified 
static response relation does not agree with the effective temperature
defined through the violation of the Einstein relation.

It would also be very interesting to extend the present results,
which pertain to non-linear response to an external potential, to 
material properties, in the spirit of the numerical investigation of Haxton 
and Liu \cite{HL}. 

The present results (and essentially all previous work on
fluctuation-dissipation relation violation in systems of interacting particles)
are concerned with a fluid undergoing a simple shear flow. A flow with a non-uniform
velocity gradient, \textit{e.g.} a Poisseuile flow, would introduce another length scale 
into the problem. Having three length scales in the problem, a microscopic length
(colloidal particle size), the characteristic length of the external potential
and a length characterizing the non-uniform flow, would make the present approach
inapplicable.

Finally, we should note that, while we derived a formal microscopic expression 
for the effective temperature, Eq. \eqref{Teff} together with Eqs. (\ref{D}-\ref{Dint})
and (\ref{mu}-\ref{muint}), we did not calculate the effective temperature theoretically.
To do this, one would have to develop a theoretical approach to calculate
interaction contributions to the self-diffusion tensor \eqref{Dint} 
and the friction tensor \eqref{muint}. In an earlier work \cite{Szamel} one of us 
derived (but did not evaluate) mode-coupling expressions for these quantities. 
A different mode-coupling approach was proposed by Kr\"uger and Fuchs \cite{KF}. 
These authors evaluated their expression for the effective temperature only in the 
vicinity of the mode-coupling transition. This fact makes a direct 
comparison of our simulational results with their predictions difficult.

We thank E. Flenner for comments on the manuscript and 
gratefully acknowledge the support of NSF Grant CHE 0909676. GS thanks
Yukawa Institute, Kyoto University,
where this work was completed, for its hospitality.


\begin{thebibliography}{99}
\bibitem{nesm} See, \textit{e.g.}, studies of the fluctuation theorem
(D.J. Evans and D.J. Searles, Adv. Phys. \textbf{51}, 1529 (2002)),
Jarzynski equality (C. Jarzynski, J. Stat. Mech.: Theor. Exp. \textbf{P09005}, (2004)),
and macroscopic fluctuation theory for non-equilibrium states (G. Jona-Lasinio,
Prog. Theor. Phys. Suppl. \textbf{184}, 262 (2010).
\bibitem{LCreview} For a recent review see L.F. Cugliandolo, arXiv: 1104.4901.
\bibitem{CKP} L.F. Cugliandolo, J. Kurchan, and L. Peliti, Phys. Rev. E
\textbf{55}, 3898 (1997).
\bibitem{BerBar} L. Berthier and J.-L. Barrat, Phys. Rev. Lett. \textbf{89},
095702 (2002); J. Chem. Phys. \textbf{116}, 6228 (2002).
\bibitem{Liu}  I. Ono \textit{et al}, Phys. Rev. Lett. \textbf{89}, 095703 (2002);
C.S. O'Hern, A.J. Liu, and S.R. Nagel,
Phys. Rev. Lett. \textbf{93}, 165702 (2004).
\bibitem{LSS} B. Lander, U. Seifert and T. Speck, Europhys. Lett. \textbf{92}, 58001 
(2010).
\bibitem{ZS} M. Zhang and G. Szamel, Phys. Rev. E \textbf{83}, 061407 (2011).
\bibitem{HS} K. Hayashi and S.-i. Sasa, Phys. Rev. E \textbf{69}, 066119 (2004).
\bibitem{Szamel} G. Szamel, Phys. Rev. Lett. \textbf{93}, 178301 (2004).
\bibitem{SS} T. Speck and U. Seifert, Europhys. Lett. \textbf{74}, 391 (2006).
\bibitem{KF} M. Kr\"uger and M. Fuchs, Phys. Rev. Lett. \textbf{102}, 135701 (2009);
Phys. Rev. E \textbf{81}, 011408 (2010); 
Prog. Theor. Physics Suppl. \textbf{184}, 172 (2010).
\bibitem{ZRA} F. Zamponi, G. Ruocco, and L. Angelani, 
Phys. Rev. E \textbf{71}, 020101(R) (2005). 
\bibitem{fluctrel} D.J. Evans, E.G.D. Cohen, and G.P. Morriss, Phys. Rev. Lett.
\textbf{71}, 2401 (1993).
\bibitem{IB} P. Ilg and J.-L. Barrat, Europhys. Lett. \textbf{79}, 26001 
(2007).
\bibitem{HL} T.K. Haxton and A.J. Liu, Phys. Rev. Lett. \textbf{99}, 195701 (2007).
\bibitem{BMW} M. Baiesi, C. Maes, and B. Wynants, Phys. Rev. Lett. \textbf{103}, 010602
(2009); J. Stat. Phys. \textbf{137}, 1094 (2009).
\bibitem{SS2} See also U. Seifert and T. Speck, Europhys. Lett. \textbf{89}, 10007 (2010).
\bibitem{SBL} G. Szamel, J. B\l awzdziewicz and J.A. Leegwater, Phys. Rev. A \textbf{45},
R2173 (1992).
\bibitem{FC} See, \textit{e.g.}, M. Fuchs and M. Cates, J. Rheol. \textbf{53}, 957 (2009).
\bibitem{GS} G. Szamel, J. Chem. Phys. {\bf 114}, 8708 (2001).
\bibitem{comment1} In practice, to improve the statistics, 
we applied this potential to 14 randomly selected particles. 
\bibitem{Resibois} For a general introduction to Chapman-Enskog expansion, 
see, \textit{e.g.},
P. R\'{e}sibois and M. de Leener, \textit{Classical Kinetic Theory of Fluids} 
(Wiley, New York, 1977).
\bibitem{Titulaer} U.M. Titulaer, Physica \textbf{91A}, 321 (1978). 
\bibitem{Titulaer2} U.M. Titulaer, Physica \textbf{100A}, 234 (1980).

\end{thebibliography}
\end{document}